\documentclass[apjl]{emulateapj}

\newcommand{\OIII}{[\mbox{O\,\textsc{iii}}]}
\newcommand{\oiii}{[\mbox{O\,\textsc{iii}}]}

\newcommand{\OI}{[\mbox{O\,\textsc{i}}]}
\newcommand{\nii}{[\mbox{N\,\textsc{ii}}]}

\newcommand{\SII}{[\mbox{S\,\textsc{ii}}]}

\newcommand{\kms}{km s$^{-1}$}
\newcommand{\Ha}{H$\alpha$}     
\newcommand{\ha}{H$\alpha$}   
\newcommand{\Hb}{H$\beta$}  
\newcommand{\hb}{H$\beta$}

\newcommand{\ergs}{erg s$^{-1}$} 
\usepackage[T1]{fontenc}
\usepackage{lmodern}
\usepackage [normalem]{ulem}
\usepackage{color}
\usepackage{hyperref}
\usepackage{natbib}
\usepackage{tikz}
\usepackage{graphicx}
\usepackage{bigstrut}
\usepackage[export]{adjustbox}
\usepackage{afterpage}
\usepackage{enumitem}
\setlist[enumerate]{itemsep=0mm}
\usetikzlibrary{positioning,fit,calc}
\usetikzlibrary{shapes,backgrounds}
\usepackage{float}
\restylefloat{table}
\newcommand{\Rnum}[1]{\uppercase\expandafter{\romannumeral #1\relax}}
\usepackage{natbib}
\begin{document}

\title{Outflow kinematics manifested by the H$\alpha$ line: gas outflows in Type 2 AGNs. IV.}
\author{Daeun Kang$^{1}$ }
\author{Jong-Hak Woo$^{1}$\altaffilmark{$\dagger$}}
\author{Hyun-Jin Bae$^{1,2}$}

\affil{$^{1}$Astronomy Program, Department of Physics and Astronomy, Seoul National University, Seoul 151-742, Republic of Korea} 
\affil{$^{2}$Department of Astronomy and Center for Galaxy Evolution Research, Yonsei University, Seoul 120-749, Korea}
\altaffiltext{$\dagger$}{Author to whom any correspondence should be addressed: woo@astro.snu.ac.kr}

\begin{abstract}
		Energetic ionized gas outflows driven by active galactic nuclei (AGN) have been studied as a key phenomenon related to AGN feedback. To probe the kinematics of the gas in the narrow line region, \oiii\ $\lambda$5007 has been utilized in a number of studies, showing non-virial kinematic properties due to AGN outflows. In this paper, we statistically investigate whether the \ha\ emission line is influenced by AGN driven outflows, by measuring the kinematic properties based on the \ha\ line profile, and by comparing them with those of \oiii. Using the spatially integrated spectra of $\sim$37,000 Type 2 AGNs at z < 0.3 selected from the SDSS DR7, we find a non-linear correlation between \ha\ velocity dispersion and stellar velocity dispersion, which reveals the presence of the non-gravitational component, especially for AGNs with a wing component in \ha. The large \ha\ velocity dispersion and velocity shift of luminous AGNs are clear evidence of AGN outflow impacts on hydrogen gas, while relatively smaller kinematic properties compared to those of \oiii\ imply that the observed outflow effect on the \ha\ line is weaker than the case of \oiii.

\end{abstract}

\section{Introduction}
\label{sec:introduction}

	The scaling relations between black hole mass and galaxy properties suggest the coevolution of black holes and galaxies \citep{Kormendy2013}, for which AGN feedback may play a crucial role (e.g.~\citealt{Somer2008, Dubois2013, DeGraf2014},  see \citealt{King2015} for a review). 
	Gas outflows related to radiative mode of AGN feedback have been regarded as one of the feedback mechanisms~\citep{Fabian2006, Ciotti2007}, since energetic gas outflows may influence star formation over galactic scales by blowing out the surrounding interstellar medium (\citealt{Dubois2013}, see review by \citealt{Fabian2012}). 
	
	The high-ionization \oiii\ $\lambda$5007 emission line has been frequently used to trace the ionized gas outflows in the narrow-line region (NLR), for investigating outflow properties using individual AGNs and a large sample \citep{Boroson2005, Greene2005, Cren2010, Bae2014, Harrison2014, Liu2014, Karou2016, Woo2016}. For example, based on a sample of $\sim$400 quasars selected from the Sloan Digital Sky Survey (SDSS), \cite{Boroson2005} suggested that both black hole mass and Eddington ratio play a role in determining the \oiii\ kinematics.  \cite{Cren2010} reported that the distributions of the host galaxy inclination are systematically different between AGNs with blue- and red-shifted \oiii, supporting that outflows are biconical and a dusty stellar disk preferentially obscures a part of the cone behind the disk. 
	
	To understand AGN-driven outflows as a potential feedback mechanism in the context of galaxy evolution, it is important to investigate how common and how energetic these outflows are and how outflows are connected to star formation. To build up a robust outflow demography, \citet[][hereafter Paper~\Rnum{1}]{Woo2016} uniformly examined the \oiii\ kinematics of $\sim$39,000 Type 2 AGNs at z $<$ 0.3 \citep[see also][]{Bae2014}. They adopted a single- or double-Gaussian function to fit the \oiii\ line profile, and measured the luminosity-weighted velocity shift and velocity dispersion of \oiii. A majority of luminous AGNs shows a broad wing component in \oiii, which represents non-gravitational kinematics, i.e., outflows. Also, they found that \oiii\ velocity dispersion is  larger than stellar velocity dispersion by an average factor of 1.3-1.4, suggesting that the relatively strong outflows, which is comparable to gravitational kinematic component, are prevalent in Type 2 AGNs. The distribution in the measured velocity--velocity dispersion diagram of \oiii\ is dramatically different from that of star forming galaxies \citep{Woo2017}, while it is well reproduced by the Monte Carlo simulations using the combined model of biconical outflows and a dusty stellar disk (e.g., \citealt{Bae2016}, hereafter Paper~\Rnum{2}). 

	While the \ha\ emission line is one of the strongest lines in the rest-frame optical range, \ha\ is less utilized compared to \oiii\ in AGN outflow studies due to a few reasons. 
	First, there is a downside of using \ha\ to trace AGN-driven outflows since \ha\ is also emitted by star-forming region.
	Thus, the total \ha\ line profile observed within an aperture (e.g., 3\arcsec\ in the case of the SDSS spectroscopy) represents a mixed nature of gas that is photoionized by AGN as well as star formation. 
	Second, in the case of Type 1 AGNs, there is an additional very broad component ($>$$\sim$1000 \kms) originated from the broad-line region (BLR), hence, it is difficult to probe the outflows in the NLR unless a high spatial resolution is available to spatially separate the NLR from the BLR or a sophisticated spectral decomposition is performed to isolate the narrow component from the very broad component \citep[e.g.,][]{Woo2014, Eun2017}.
		
	Although the significance of the outflow kinematics manifested by \ha\ may be smaller than that by \oiii, \ha\ can provide valuable constraints to study non-gravitational component, i.e., outflows as well as virial component, i.e., rotation in the gas kinematics. \cite{Bae2014} compared the kinematics of \ha\ and \oiii\ in Type 2 AGNs, reporting that the fraction of AGNs with outflow signatures based on the \ha\ velocity shift with respect to systemic velocity is smaller than that based on the \oiii\ velocity shift, and that the distributions of the \ha\ velocity shift are similar between AGNs and star-forming galaxies. 
    Nevertheless, since \cite{Bae2014} used the peak of the \Ha\ line profile, rather than the flux-weighted center (the first moment) of the line, their measured velocity shift does not fully represent the outflow velocity, since the peak of the line is often strongly influenced by the gravitational (rotational) component rather than the non-gravitational outflow component \citep[e.g.,][]{Karou2016}. 
	
	In this paper, as the fourth of a series of papers on AGN outflows, we investigate the gas kinematics traced by the \ha\ line, by measuring the first and second moments of the \ha\ line profile for calculating velocity and velocity dispersion. We use the sample of Type 2 AGNs from Paper~\Rnum{1}, which selected Type 2 AGNs from the SDSS DR7 and reported the demography of AGN outflows and the kinematics of \oiii\ in detail. 
        We describe how we measure the \ha\ kinematics in Section~\ref{sec:sample_method}. In Section~\ref{sec:results}, 
    we present the properties of the \ha\ kinematics, and compare them with those of \oiii. We discuss the results and their implications in Section~\ref{sec:discussion}. Conclusions and summary are given in Section~\ref{sec:summary and conclusion}. In this paper, we adopted $\Lambda$CDM cosmology with cosmological parameters: H$_{0}$ = 70~km s$^{-1}$ Mpc$^{-1}$, $\Omega_{m}$ = 0.30, and $\Omega_{\Lambda}$ = 0.70.
	
\section{Sample \& Methodology}
\label{sec:sample_method}

\subsection{Sample selection}

    In order to probe the kinematics of \ha\ emission line in this paper, we use the same sample of
$\sim$39,000 Type 2 AGNs at z $<0.3$, which were used for the detailed study of the ionized gas outflows based on \oiii\ (Paper~\Rnum{1}). The details of the sample selection is presented in Paper~\Rnum{1} and \cite{Woo2017} \citep[see also][]{Bae2014}. 
    Here, we briefly summarize the selection criteria. For statistical studies of gas outflows in Type 2
AGNs, we selected galaxies with well-defined emission lines, i.e.,  signal-to-noise (S/N) ratio $>$ 3 for four major emission lines, \ha, \oiii\ $\lambda$5007, \Hb, and \nii\ $\lambda$6584, the amplitude-to-noise ratio $>$ 5 for \ha\ and \oiii, and S/N $>$ 10 for continuum, using the SDSS Data Release 7~\citep{Aba2009}. 

    Each galaxy is classified as Seyfert galaxies, low-ionization nuclear emission-line region (LINER),
composite objects or star forming galaxies based on the emission line flux ratios~\citep{Kauff2003, Kewley2006}. 
We use a loose criterion of \oiii/\Hb\ > 3 to distinguish Seyfert galaxies from less energetic AGNs, i.e., LINERs. Since the separation between Seyfert galaxies and LINERs is only based on the \oiii/\Hb\ ratio, it is possible that more luminous and higher Eddington ratio AGNs than typical LINERs are included in the LINER group. Actually, 3.9\% of LINERs have  high \oiii\ luminosity (i.e.,  $>$ 10$^{41}$ \ergs), while 14.4\% of LINERs have high Eddington ratio (i.e., $>$ 0.01). These results are
similar to the study by \cite{Kewley2006}, who reported that 92$\%$ of their LINERs, which were classified by \oiii/\Hb\ < 3, belong to the more strictly defined
LINER group based  both \SII/\ha\ and \OI/\ha\ ratios. Note that these LINERs with relatively high luminosity and high Eddington ratio do not significantly
change the main results (see Section 3). Thus, we use a loose definition of LINERs in this paper, in order to separate less energetic AGNs from Seyfert galaxies.
	
    To properly measure the kinematics of the \ha\ emission line, we exclude a total of 1,681 objects from the sample. 
First, we eliminate Type 1 AGNs with a very broad component in the \Ha\ line profile, in order to focus on Type 2 AGNs. Some of these objects were confirmed as Type 1 AGNs and in detail studied by \cite{Woo2015} and \cite{Eun2017}. In addition to these objects, we conservatively exclude additional Type 1 AGN candidates by carefully investigating the profile of \Ha\ and checking whether the measured velocity dispersion of \Ha\ is abnormally high (i.e., $>$ 700~\kms). It is possible that narrow-line Seyfert 1 galaxies might be included in Seyfert galaxies or LINERs, while we tried to identify and exclude type 1 AGNs from the sample \citep[see][]{Eun2017}. Nevertheless, since the number of AGNs with very high \ha\ velocity dispersion is extremely small, the effect of the potential contamination of Type 1 AGNs is insignificant. Second, we exclude 1,197 objects due to the lack of suitable stellar velocity dispersion measurements or too small velocity dispersion of the \ha\ line (i.e., velocity dispersion of \ha\ is smaller than 30~\kms). As a result, we finalize a total sample of 37,301 Type 2 AGNs for the \Ha\ study (see Table~\ref{table:sample}).

 \renewcommand{\arraystretch}{1.25}
 \begin{deluxetable}{cccc}[t]
 	\tablecolumns{4} 
 	\tablecaption{The number of finally used objects \label{table:sample}}
	
	\tablehead
	{	\colhead{redshift} &
		\colhead{composite objects} &
		\colhead{LINER} &
		\colhead{Seyfert}
	}
	\startdata
	z < 0.05 		& 3570 (1967) 		& 3429 (1029) 	& 2142 (905)\\
	0.05 < z < 0.1 	& 7520 (4666) 		& 4356 (1890) 	& 4456 (2636)\\
	0.1 < z < 0.2	& 4396 (3006) 		& 2478 (1430) 	& 4264 (2610) \\ 
	0.2 < z < 0.3	&   311 (203) 		& 170 (84) 	& 209 (116)
 	\enddata
	\tablecomments{The number of sample at each redshift range for each group: Seyfert galaxies, LINERs and composite objects. The number of sample with double Gaussian profile is presented inside the bracket.}
 \end{deluxetable}

\subsection{Methodology}

	Adopting the same method of measuring gas kinematics as we used for the \oiii\ line in Paper~\Rnum{1}, 
we measure the velocity shift and velocity dispersion of \Ha\ to trace the kinematics of \ha\ gas, by fitting \ha\ and \nii\ doublet ($\lambda$6549, $\lambda$6583) with the MPFIT package~\citep{Mark2009}. We choose a double Gaussian model to fit each emission line. However,
if the amplitude-to-noise ratio of the second Gaussian component is less than 3, we instead use a single Gaussian model as similarly done for \oiii\ in Paper~\Rnum{1}. Note that since we conservatively determine whether a wing component is required for the fit, the fraction of AGNs with a double Gaussian profile in \Ha\ should be considered as a lower limit. 

Based on the best-fit model, we calculate the 1st and 2nd moment as:
	 \begin{eqnarray}
	\lambda_{0} = {\int \lambda f_\lambda d\lambda \over \int f_\lambda d\lambda}
	\label{eq:fmom}
\end{eqnarray}

\begin{eqnarray}
	[\Delta\lambda_{H\alpha}]^2 = {\int \lambda^2 f_\lambda d\lambda \over \int f_\lambda d\lambda} - \lambda_0^2
	\label{eq:smom}
\end{eqnarray}
	 to obtain velocity and velocity dispersion of \Ha. Then we calculate the velocity shift of \ha\ with respect to the systemic velocity measured from stellar absorption lines during the continuum subtraction process (see Paper~\Rnum{1}).
To determine the measurement uncertainty of each parameter (i.e., flux, velocity shift and velocity dispersion of \ha), we perform
Monte Carlo simulations by generating 100 mock spectra for each object using flux errors. Then, we adopt the standard deviation of the distribution of the measurements as an 1$\sigma$ uncertainty. 

For AGN energetics, we adopt bolometric luminosity, black hole mass and Eddington luminosity as discussed in Paper~\Rnum{1}. To obtain bolometric luminosity, we use dust extinction uncorrected luminosity of the \oiii\ emission line, multiplying it by bolometric correction, 3500~\citep{Heck2004} while black hole mass is determined based on the relationship between black hole mass and stellar mass~\citep{Marconi2003}.
Note that AGN luminosity and black hole mass have large uncertainties since they are not direct measurements. Nevertheless, they are good enough to 
investigate the relative trend between outflow kinematics and AGN energetics.

\section{Results}
\label{sec:results}

\subsection{\Ha\ Luminosity}

\begin{figure}[t]
	\center
	\includegraphics[width=0.48\textwidth]{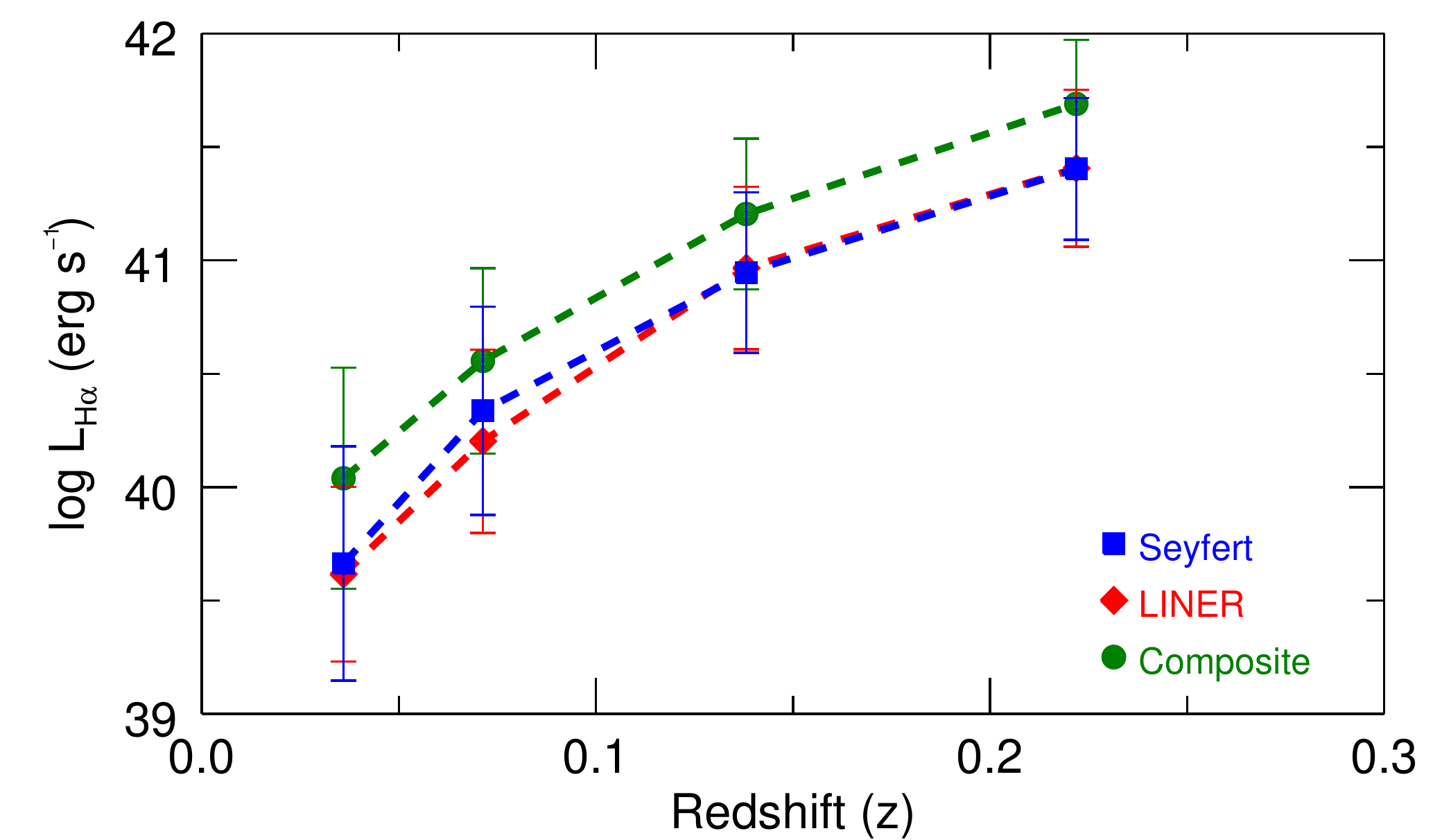}
	\caption{Mean \Ha\ luminosity of galaxies in each group as a function of redshift. \label{fig:zlum}}
\end{figure}

	In this section, we compare the luminosities of the \oiii\ and \ha\ emission lines. Figure~\ref{fig:zlum} presents the mean \Ha\ luminosity of each group as a function of redshift. We find that \Ha\ luminosity increases as a function of redshift for all three groups, as similarly found in the case of \oiii\ (see Figure 2 in Paper~\Rnum{1}), which reflects the fact that more luminous AGNs were observed at higher redshift due to the selection limit. 
	Interestingly, the mean \ha\ luminosity of composite objects is higher by an average factor of $\sim$1.7 than pure AGNs (Seyfert galaxies and LINERs) in the overall redshift range, while the \oiii\ luminosity of pure AGNs was higher by an average factor of $\sim2.9$ than that of composite objects as shown in Paper I. This is due to the fact that the \OIII\ emission mainly comes from AGN, while \Ha\ can be emitted from star forming region as well as AGN~\citep{Gar1996}. Thus, this trend suggests that \ha\ luminosity is not a good proxy for AGN luminosity unless the contribution from star forming region is properly corrected for.

\begin{figure}%[t]
	\center
	\includegraphics[width=0.48\textwidth]{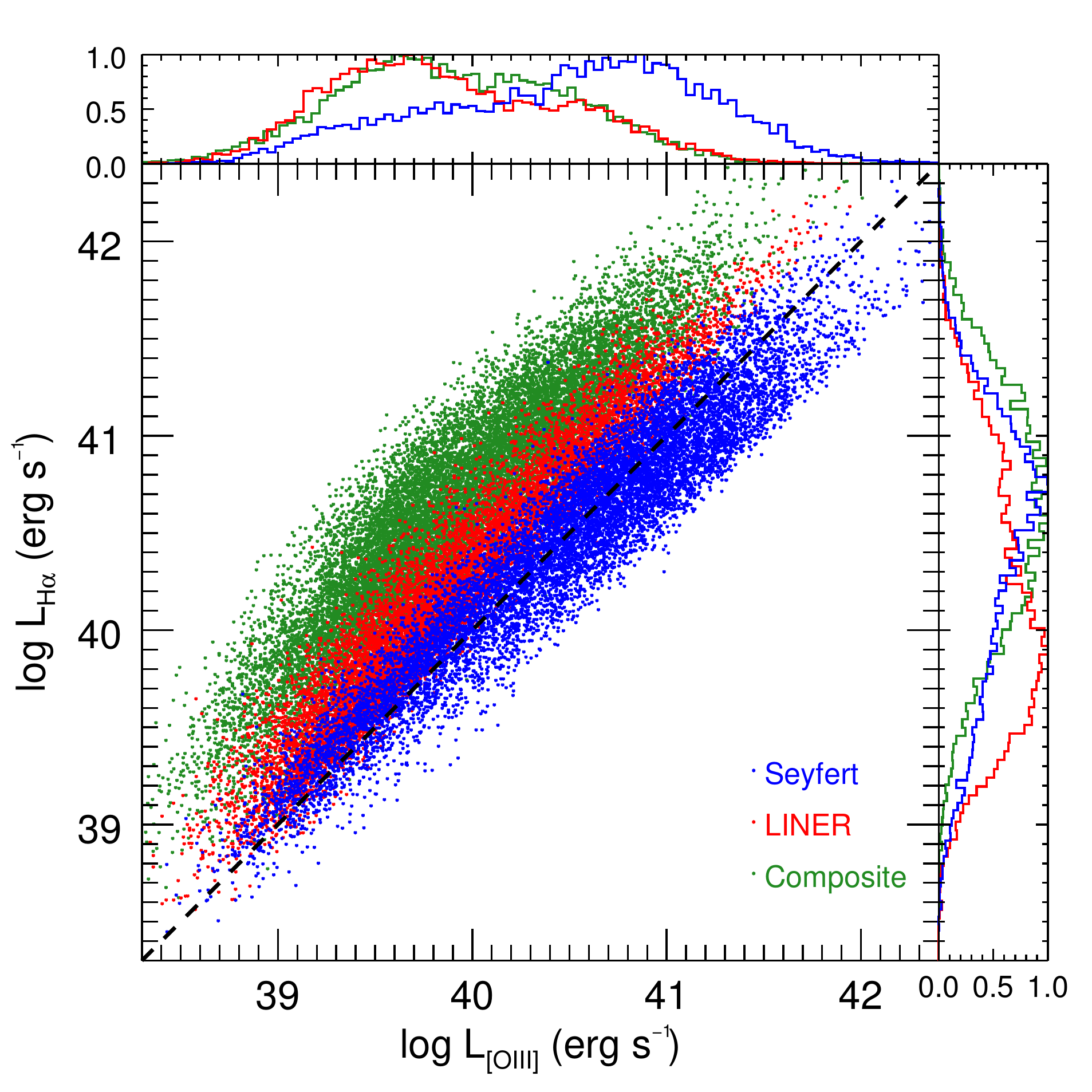}
	\caption{Comparison of \ha\ luminosity with \oiii\ luminosity. Both are dust extinction uncorrected. Color represents each group : Seyfert galaxies (blue), LINERs (red), and composite objects (green).  \label{fig:o3ha1}}
\end{figure}

\begin{figure*}
	\center
	\includegraphics[width=0.97\textwidth]{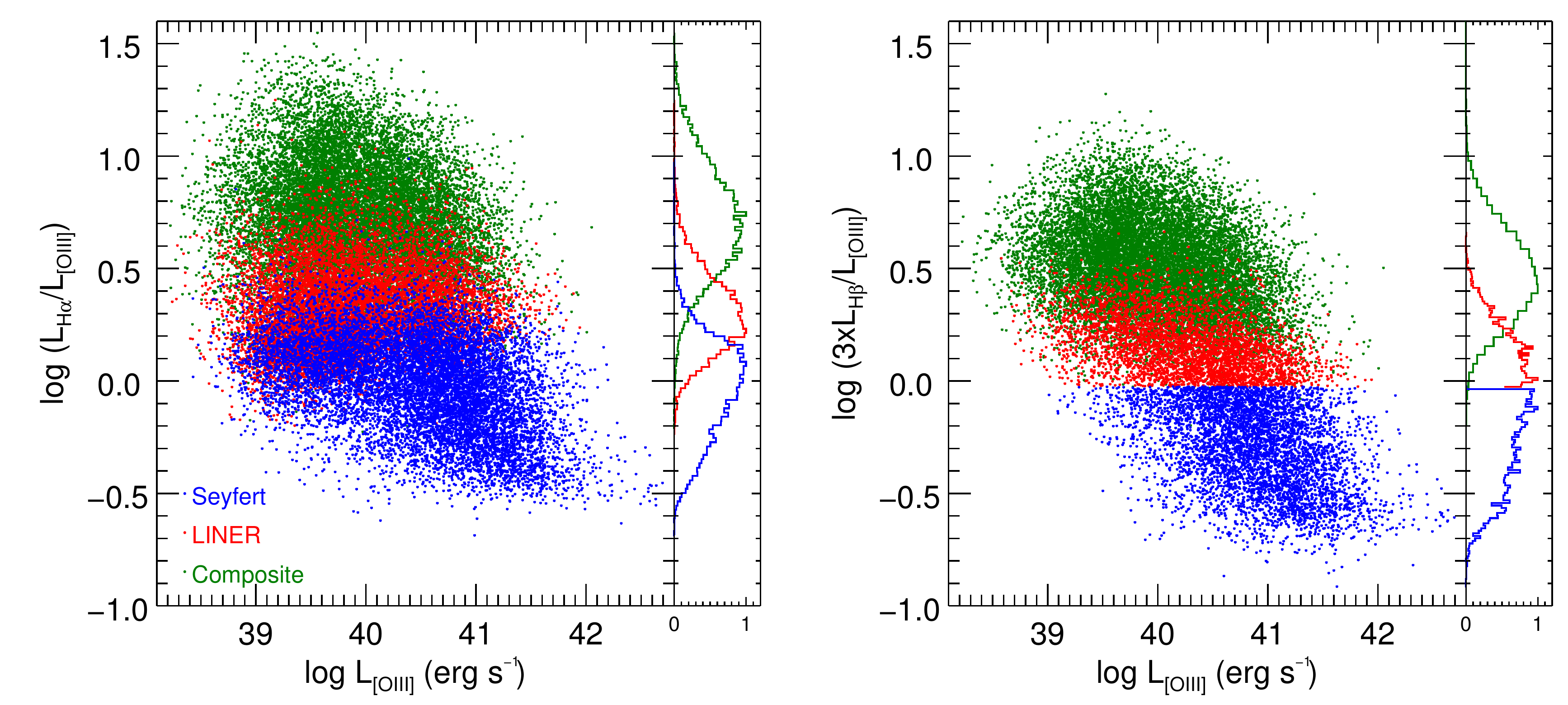}
	\caption{Comparison of \Ha-to-\oiii\ luminosity ratio with \oiii\ luminosity. Color code is same as Figure~\ref{fig:o3ha1}. On the right panel, we corrected \ha\ luminosity using \Hb\ luminosity to compensate the difference of dust extinction between \ha\ and \oiii. \label{fig:o3ha2}} 
\end{figure*}

\begin{figure*}[t]
	\center
	\includegraphics[width=0.98\textwidth]{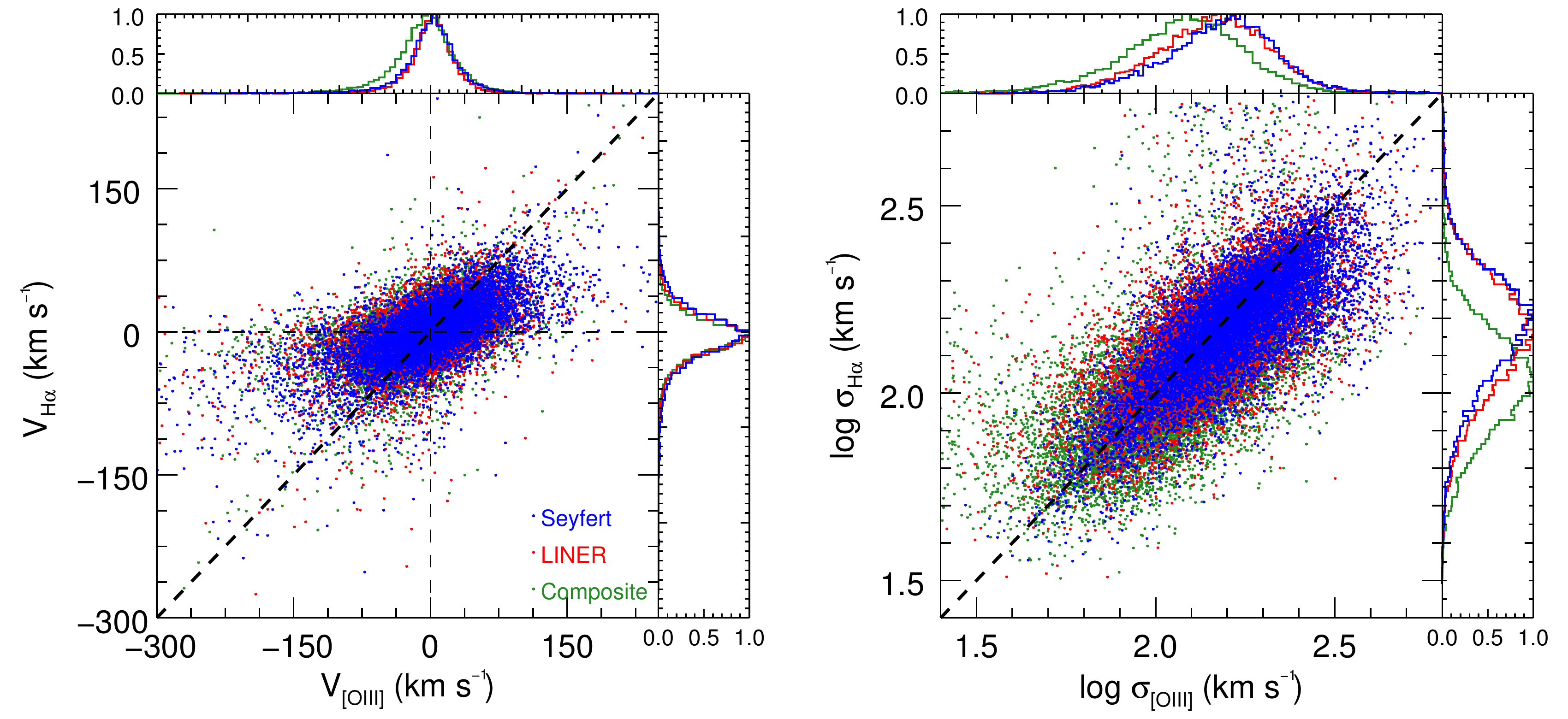}
	\caption{Comparison of velocity shift and velocity dispersion of \ha\ with \oiii. Color code is same as Figure~\ref{fig:o3ha1}.  		\label{fig:o3ha3}} 
\end{figure*}

In Figure~\ref{fig:o3ha1}, we directly compare \ha\ and \oiii\ luminosities. While \ha\ luminosity broadly correlates with \oiii\ luminosity, \ha\ luminosity is on average higher than that of \oiii, particularly for composite objects. 
Note that these luminosities are not corrected for dust extinction and that \oiii\ suffers more extinction than \ha. The average of the \ha-to-\hb\ flux ratio is $\sim$5, which is larger than $\sim$3 expected from the Case B recombination, clearly showing the different amount of dust extinction between the two spectral ranges. If we assume that the \ha-to-\hb\ flux ratio should follow the prediction based on the Case B recombination, \Hb\ luminosity is underestimated by an average $\sim$0.22 dex. The fraction of Seyfert galaxies with \oiii\ luminosity higher than \ha\ luminosity is $\sim$50$\%$ while for the majority of composite objects \ha\ luminosity is higher than \oiii\ luminosity. 

We find that the distribution of \ha\ luminosity is similar between Seyfert galaxies and composite objects while the LINERs have an order of magnitude lower mean \Ha\ luminosity. This trend is very different from that of \OIII, which shows a similar distribution between LINERs and composite objects while Seyfert galaxies have a much higher mean \OIII\ luminosity. The difference of the luminosity distributions between \Ha\ and \oiii\ suggests that the contribution from the star forming region to the observed \Ha\ is systematically different among Seyfert galaxies, composite objects, and LINERs.
    Note that since we loosely defined LINERs, there are sources with high \OIII\ luminosity 
(i.e., \oiii\ luminosity $>$ 10$^{41}$ \ergs) and high Eddington ratios (i.e., $>$ 0.01) in LINER group. However, even if we re-classify these AGNs as Seyfert galaxies rather than LINERs, the overall shape and difference of the luminosity distributions of Seyfert galaxies and LINERs 
do not significantly change.
	
To further investigate the relation between \ha\ and \oiii\ luminosities, we present the \ha-to-\oiii\ luminoisty ratio as a function of \oiii\ luminosity in Figure~\ref{fig:o3ha2} (left panel). Since \oiii\ suffers more extinction than \ha, we instead use \Hb\ luminosity rather than the \Ha\ luminoisty, after multiplying a factor of three. Thus, we present the extinction-independent \Ha-to-\oiii\ ratio (Figure~\ref{fig:o3ha2}, right panel). 
Considering the fact that the \Hb\ line is much weaker than \Ha, we avoid uncertain \Hb\ luminosity measurements by excluding objects with a large fractional error, i.e., $>$ 1$\sigma$ in \hb\ luminosity.
Consequently, 90$\%$ of composite objects, 40$\%$ of LINERs and 47$\%$ of Seyfert galaxies remained.If \ha\ luminosity linearly correlates with \oiii\ luminosity, Figure~\ref{fig:o3ha2} ought to show a flat trend. Instead, the luminosity ratio decreases, suggesting that the contribution from star forming region to \Ha\ is significant in low luminosity AGNs and particularly in composite objects. Note that the \hb-to-\oiii\ ratio in the NLR also varies depending on the gas and ionization properties. However, it is shown that there is a systematic difference
of the \Ha-to-\oiii\ luminosity ratio, depending on the contribution from star-forming region.

\subsection{\Ha\ Kinematics}

In this section, we investigate the kinematics traced by the \ha\ line and compare them with those of \oiii\ and stellar lines.
First, we present  the velocity and velocity dispersion of \ha\ and \oiii\ in Figure~\ref{fig:o3ha3}. 
As discussed in Paper~\Rnum{1}, the velocity shift is relatively small for both \OIII\ and \Ha\ due to the fact that the direction of AGN outflows is highly inclined from the line-of-sight (i.e., Type 2 AGNs). Despite of the orientation effect, there are AGNs with a relatively large velocity shift, for which the amplitude of \oiii\ velocity shift is usually larger than that of \ha. 

When we compare velocity dispersion of \ha\ and \oiii, the correlation between them is tighter for Seyfert galaxies than composite objects. When we perform forward regression, the correlation slope slightly decreases from 1.01$\pm$0.01 for composite objects to 0.82$\pm$0.01 for Seyfert galaxies, and scatter also decreases from 0.11 to 0.08 dex. This is due to the fact that velocity dispersion of \oiii\ is mainly broadened by AGN activity, while \ha\ velocity dispersion is influenced by AGN as well as star forming region, which effectively reduces the total line broadening due to the lack of high velocity gas in star forming region. Note that composite objects show significantly different distribution of \ha\ velocity dispersion distribution compared to pure AGNs. We interpret that this trend is due to the contamination from star forming region, which narrows the observed total \ha\ profile. Since the kinematic properties of \ha\ correlate with those of \oiii\ in Figure~\ref{fig:o3ha3}, we expect that \ha\ kinematics will provide similar outflow properties compared to \oiii\ kinematics studied in Paper~\Rnum{1}.

\begin{figure}[t]
    \center
    \includegraphics[width=.47\textwidth]{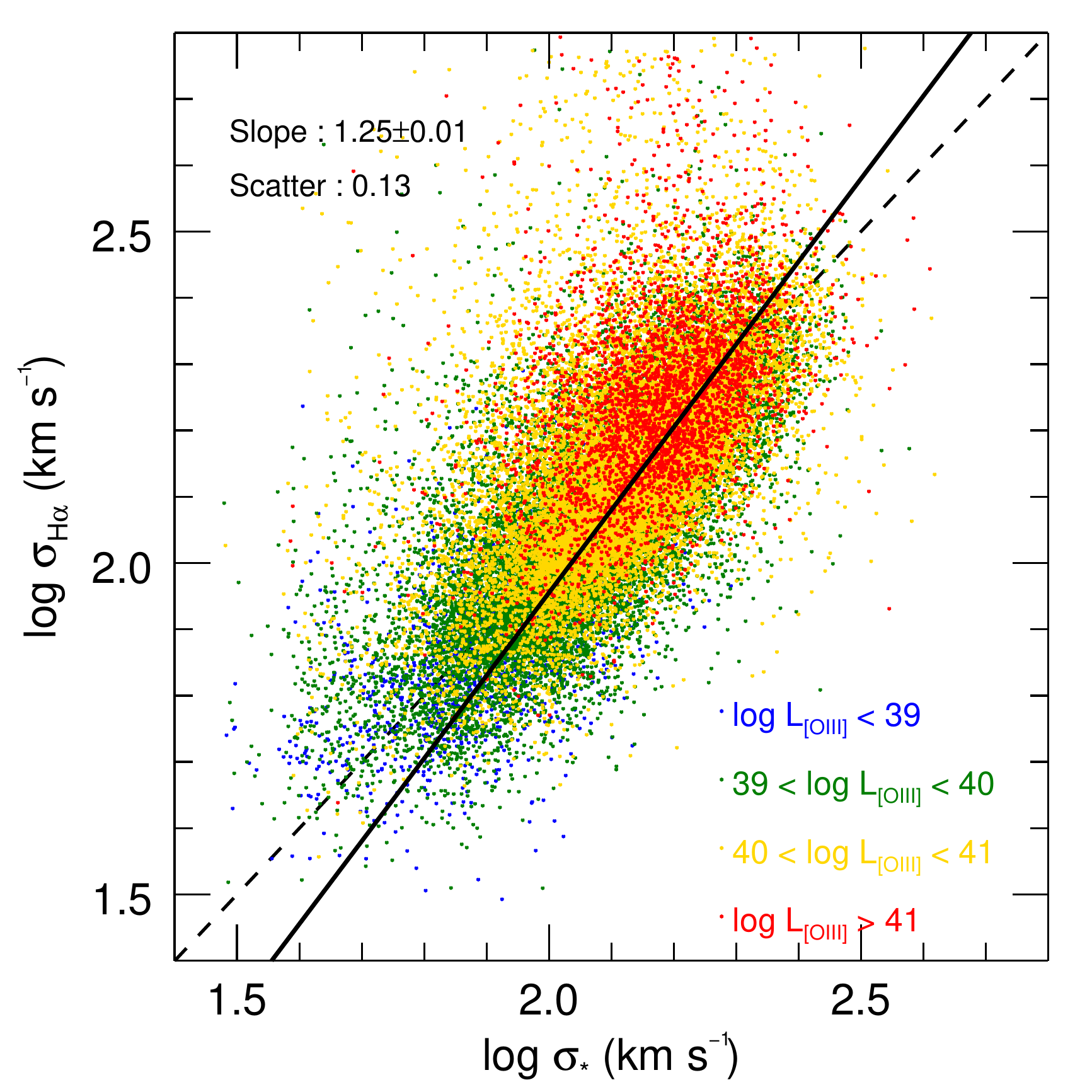}
    \caption{Comparison of \Ha\ velocity dispersion with stellar velocity dispersion for whole sample. Color represents each \oiii\ luminosity range of the sample. Solid line indicates best-fit slope and dashed line indicates one-to-one relation. \label{fig:disptotal}}
\end{figure}

Second, we investigate whether \ha\ kinematics are governed by the gravitational potential of the host galaxy or any additional component (i.e., non-gravitational component) exists, similar to the case of \oiii, by comparing \ha\ velocity dispersion ($\sigma_{H\alpha}$) with stellar velocity dispersion ($\sigma_{*}$). In Paper~\Rnum{1}, it is reported that \oiii\ velocity dispersion has a non-linear correlation with stellar velocity dispersion. While the kinematics of \oiii\ is partly governed by the gravitational potential of host galaxy, the majority of AGNs show an additional non-virial component due to AGN outflows. We expect similar results for \ha\ gas kinematics, considering the correlation between \oiii\ and \ha\ velocity dispersion in Figure~\ref{fig:o3ha3}.

Figure~\ref{fig:disptotal} compares \ha\ velocity dispersion with stellar velocity dispersion. We perform forward regression including errors of both \ha\ and stellar velocity dispersions, obtaining a slope of 1.25$\pm$0.01 with a 0.13 dex scatter. In comparison, \oiii\ showed a steeper slope, 1.43$\pm$0.01, and a larger scatter, 0.19 dex (Paper~\Rnum{1}), demonstrating that the effect of the non-virial motion is weaker in \Ha\ than \oiii. Note that for this comparison we exclude objects with unreliably high stellar velocity dispersion above 420~\kms, as recommended in the SDSS catalog. Considering the instrumental resolution of SDSS, $\sim$70~\kms, we also exclude objects of which either stellar velocity dispersion or \ha\ velocity dispersion is below 30~\kms.

\begin{figure}[t]
    \center
    \includegraphics[width=.48\textwidth]{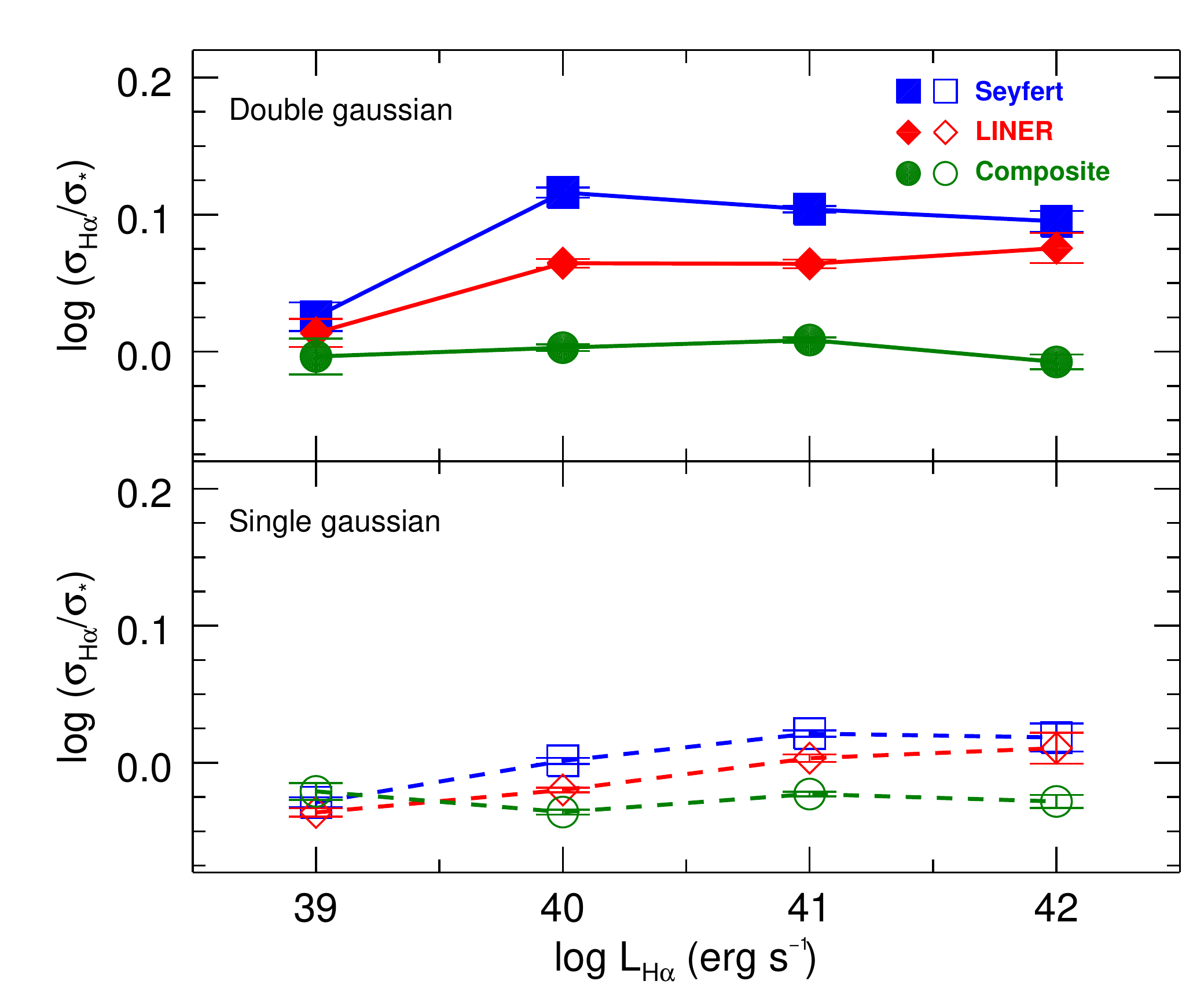}
    \caption{Mean ratio of \Ha-to-stellar velocity dispersion as a function of \ha\ luminosity for galaxies with single Gaussian \Ha\ (lower panel) and double Gaussian \Ha\ (upper panel) in each group - Seyfert galaxies, LINERs, and composite objects. Error bars represent standard errors. \label{fig:dr_lum}}
\end{figure}

    When we divide the sample in to two classes: those with a wing component in the \ha\ line profile 
(hereafter double Gaussian \ha) and those without (hereafter, single Gaussian \ha), we obtain the best-fit slope of 1.18$\pm$0.01 with scatter of 0.09 dex for single Gaussian \ha, while the best-fit slope for double Gaussian \ha\ was 1.39$\pm$0.01 with scatter of 0.16 dex. The best-fit slope for single Gaussian \ha\ is similar to that of single Gaussian \oiii\ profile, which is 1.18$\pm$0.01 (Paper~\Rnum{1}), while the best-fit slope for double Gaussian \ha\ is smaller than that of double Gaussian \oiii, which is 1.66$\pm$0.01. When we perform forward regression for Seyfert galaxies, we obtain the best-fit slope of 1.27$\pm$0.01 with 0.14 dex scatter, while the best-fit slope between \oiii\ and stellar velocity dispersion is much higher (1.69$\pm$0.01).
    The non-linear correlation between \ha\ and stellar velocity dispersions, similar to the case of \oiii\ (Paper~\Rnum{1}),
implies that there exists non-gravitational effect which makes \ha\ velocity dispersion broader, while the non-virial effect on the hydrogen line is weaker than that on \oiii. In contrast, the kinematics represented by single Gaussian \ha\ and \oiii\ is mainly governed by the gravitational potential of the host galaxy~(Paper~\Rnum{1}, \citealt{Karou2016}). 

\begin{figure}
    \center
    \includegraphics[width=.48\textwidth]{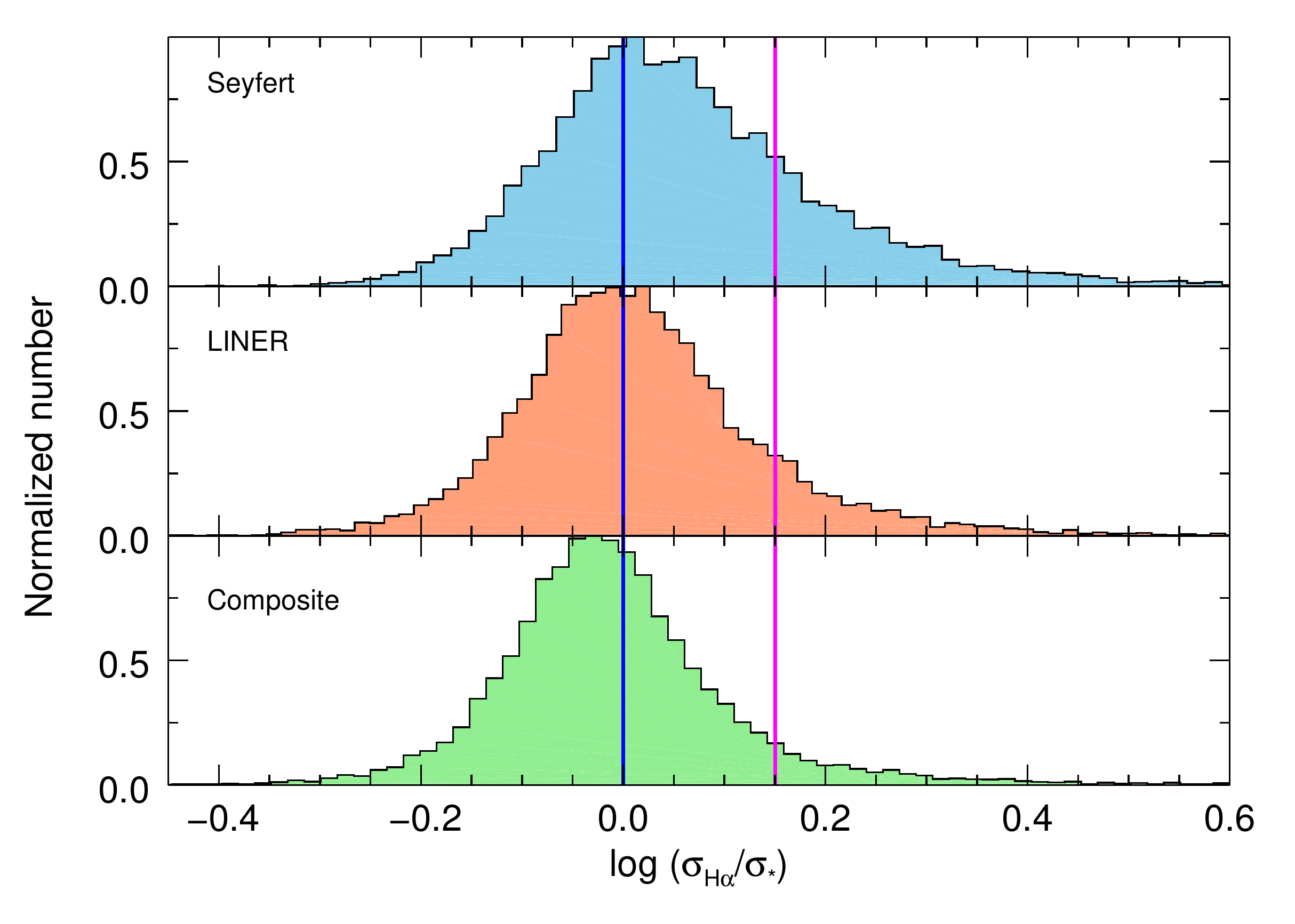}
    \caption{Histogram of \ha\ to stellar velocity dispersion for each group. Stellar velocity dispersion represents the gravitational dispersion ($\sigma_{gr}$). Vertical blue line indicates where $\sigma_{H\alpha}$ = $\sigma_{*}$ and pink vertical line indicates where $\sigma_{non-gr} = \sigma_{*}$.\label{fig:dhis}}
\end{figure}

\begin{figure*}[t]
	\center
	\includegraphics[width=0.99\textwidth, height=0.7\textwidth]{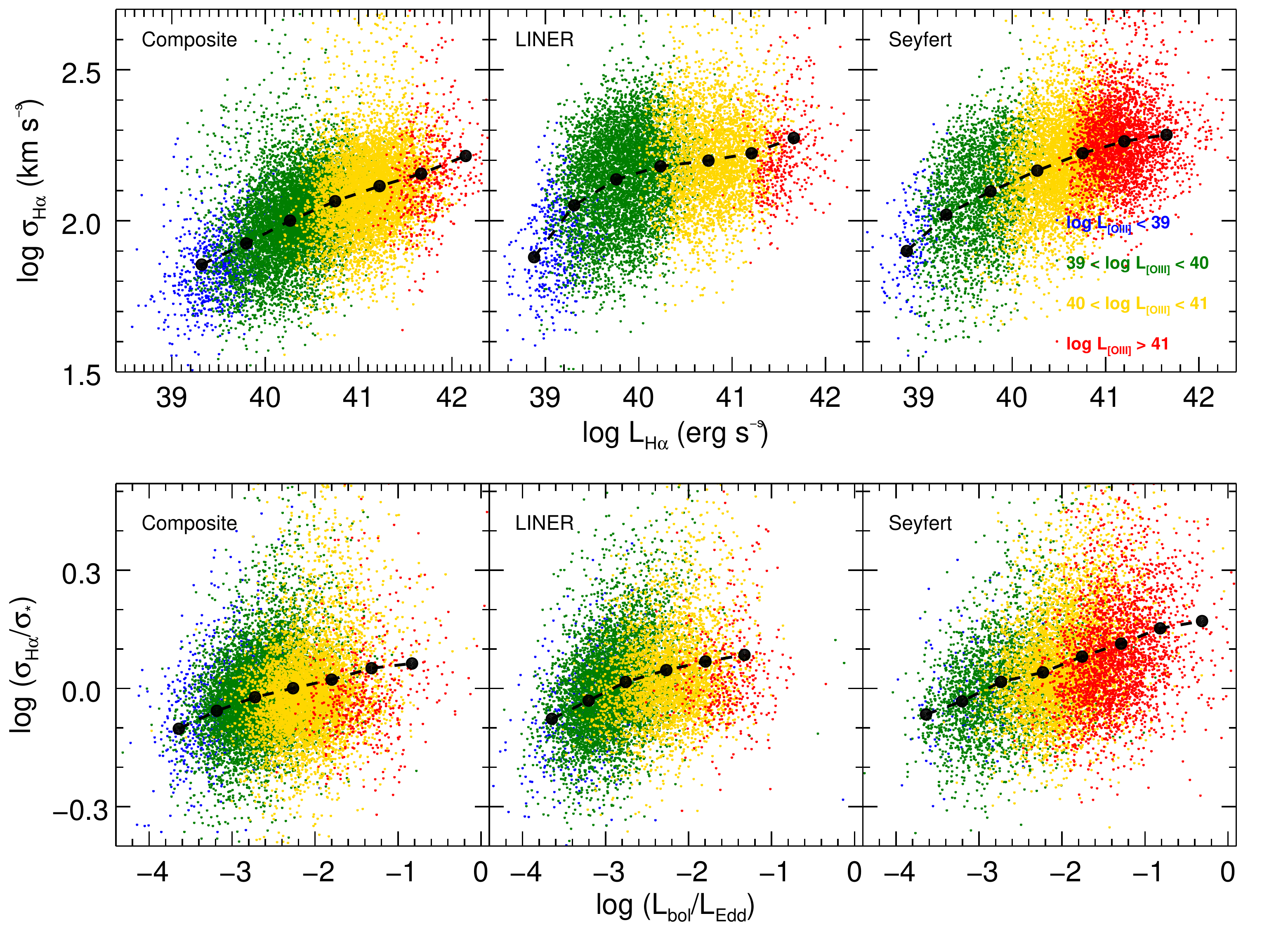}
	\caption{Comparison of \Ha\ velocity dispersion with \Ha\ luminosity (upper panel) and comparison of Eddington ratio with \ha\ to stellar velocity ratio (lower panel) for each 3 group: composite objects (left), LINERs (middle), and Seyfert galaxies (right). Color-code is same as Figure~\ref{fig:disptotal}. Filled circles represent average \ha\ velocity dispersion in each \ha\ luminosity bin (top) and average \ha\ to stellar velocity dispersion in each Eddington ratio bin (bottom). \label{fig:d_lumedd}}
\end{figure*}

To estimate the non-gravitational component in \ha, we divide \ha\ velocity dispersion by two components:
\begin{eqnarray}
	(\sigma_{H\alpha})^{2}&=(\sigma_{gr})^{2}+(\sigma_{non-gr})^{2}
\end{eqnarray}
where $\sigma_{non-gr}$ is non-gravitational component and $\sigma_{gr}$ is gravitational component.  We utilize the stellar velocity dispersion as a proxy for the gravitational component ($\sigma_{gr}$ = $\sigma_{*}$). 
For AGNs with $\sigma_{H\alpha}$ > $\sigma_{*}$, we find the \ha\ velocity dispersion is larger than $\sigma_{*}$ by an average factor of
1.35.
 
In Figure~\ref{fig:dr_lum}, we normalize \ha\ velocity dispersion by stellar velocity dispersion (hereafter dispersion ratio) and compare it with \ha\ luminosity to investigate whether the relative amount of non-gravitational effect on \ha\ kinematics is related to the emission line luminosity. Mean dispersion ratio for pure AGNs with double Gaussian \ha\ profile increases from 0.01 to 0.1, while that for pure AGNs with single Gaussian \ha\ shows nearly flat trend around 0. The trend of pure AGNs with double Gaussian \ha\ is similar to the case of \oiii, though the mean dispersion ratio value is quite smaller than that of \oiii\ since In Paper~\Rnum{1} reported that the non-gravitational component of \oiii\ was comparable to stellar velocity dispersion. On the other hand, mean dispersion ratio of composite objects with both single and double \ha\ profile does not increase as a function of \ha\ luminosity . This implies that \ha\ velocity dispersion of composite objects is dominated by virial motion.

In Figure~\ref{fig:dhis}, we present the distribution of the \ha\ to stellar velocity dispersion ratio. A vertical blue line indicates where \ha\ dispersion is equal to the stellar velocity dispersion and a vertical pink line indicates where non-gravitational kinematic component is comparable to the stellar velocity dispersion. The fraction of objects with detectable outflows (i.e., ratio over the blue line) is 65, 52, and 41\%, respectively for Seyfert galaxies, LINERs, composite objects, and the fraction of objects with a dominant non-gravitational component (i.e., ratio over the pink line) is 23, 13, and 8\%, respectively for each group. These results show that strong outflows are prevalent in pure AGNs.

Last, we discuss the measured \ha\ velocity shift. The number ratio of AGNs with a blueshifted \Ha\ to AGNs with redshifted \Ha\ is 0.88. Considering that the average error of \ha\ velocity shift measured by Monte Carlo simulation is 13.9~\kms~(1$\sigma$), when we only consider objects with reliable velocity shift measurements (i.e., \ha\ velocity shift $>$ 3$\sigma$), we obtain the ratio 1.17. In the case of double Gaussian \ha, the number ratio of blueshifted-to-redshifted \ha\ is 1.62, reflecting that blue wings are more frequently observed than red wings. Moreover, the number ratio tends to increase as a function of \ha\ luminosity, from 0.60 to 1.94 over the 4 orders of magnitude in \ha\ luminosity. A similar trend is reported for \oiii\ that \oiii\ is more frequently blueshifted than redshifted (Paper~\Rnum{1}, \citealt{Karou2016}).

\begin{figure*}[t]
    \center
    \includegraphics[width=.96\textwidth, height=.524\textwidth]{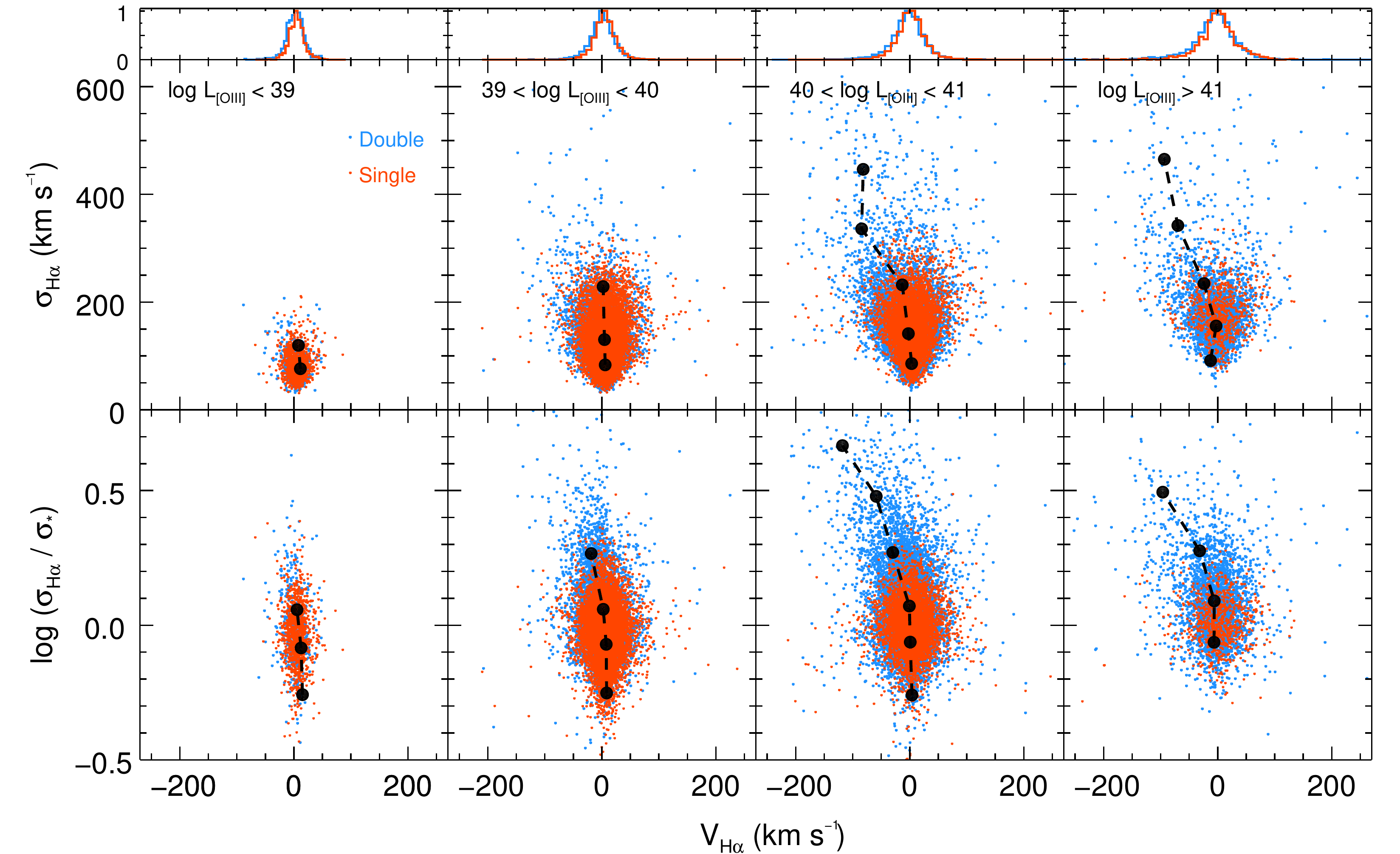}
    \caption{VVD diagram (upper panel) and comparison of \ha\ to stellar velocity dispersion with \ha\ velocity shift (lower panel) in each \oiii\ luminosity range for single Gaussian \ha\ (red) and for double Gaussian \ha\ (blue). Filled circles are mean values of \ha\ velocity shift in each range of y-axis variable where we only consider sample with reliable \ha\ velocity shift amplitude (|V| > 3$\sigma$). .\label{fig:vvd}}
\end{figure*}

\subsection{\Ha\ kinematics vs. AGN energetics}

To investigate how \ha\ kinematics are related to AGN energetics, we first compare the luminosity and velocity dispersion of the \ha\ lline in Figure~\ref{fig:d_lumedd}. \ha\ velocity dispersion increases with \ha\ luminosity albeit with a large scatter. Each group shows similar increments in \ha\ velocity dispersion as the average velocity dispersion increases by more than a factor of 2 over 4 orders of magnitude in \ha\ luminosity. As similarly reported for the \oiii\ line (Paper~\Rnum{1}), these results suggest that outflows are directly connected to AGN energetics.

Second, we compare Eddington ratio with the normalized \ha\ velocity dispersion by stellar velocity dispersion (bottom panel in Figure~\ref{fig:d_lumedd}). The normalized \ha\ velocity dispersion tends to increase with Eddington ratio, suggesting that the outflow effect on \ha\ is stronger for AGNs with higher Eddington ratio. The mean of the normalized \ha\ velocity dispersion increases by a factor of $\sim$1.8 over 3 orders of magnitude in Eddington ratio, which is similar to the case of \oiii\ presented in Paper~\Rnum{1}.
	
Third, we investigate the velocity-velocity dispersion (hereafter VVD) diagram of \ha, by plotting AGNs with single Gaussian \ha\ (red) and double Gaussian \ha\ (blue) in each \oiii\ luminosity bin (Figure~\ref{fig:vvd}). In general, AGNs with strong outflows (i.e., high velocity dispersion and/or high |V$_{H\alpha}$|) preferentially appear in high \oiii\ luminosity range. AGNs with double Gaussian \ha\ are dominant among AGNs with strong outflow kinematics with velocity shift up to $\sim$200 \kms and velocity dispersion up to $\sim$600 \kms, while the velocity shift and dispersion are limited to small range in the case of AGNs with single Gaussian \Ha. 

Using only reliable velocity measurements (|V| > 3$\sigma$), we present the average velocity shift in each velocity dispersion bin (filled circles) in Figure~\ref{fig:vvd}, which shows that \oiii\ luminous AGNs typically have large blue shift in \ha\ although not all luminous AGNs have high velocity shift values. In contrast, the majority of low-luminosity AGNs have small velocity shift close to 0 \kms. 
	 
These results indicate that luminous AGNs tend to have strong outflows, which manifest a stronger non-virial kinematic component than virial kinematic component, while in the case of low-luminosity AGNs, the outflow component is often diluted by the virial component, leading to zero or small velocity shift and
small normalized velocity dispersion. We find that the VVD distribution of \ha\ is qualitatively similar to that of \oiii, and that 
the observed VVD distributions of \ha\ and \oiii\ are consistent with the interpretation that AGN outflows are biconical with a presumably large opening angle as constrained based on the Monte Carlo simulations of the VVD distribution in Paper~\Rnum{2}. 

\subsection{Outflow fractions}
\label{subsec:fraction}

\begin{figure}
	\center
	\includegraphics[width=.35\textwidth]{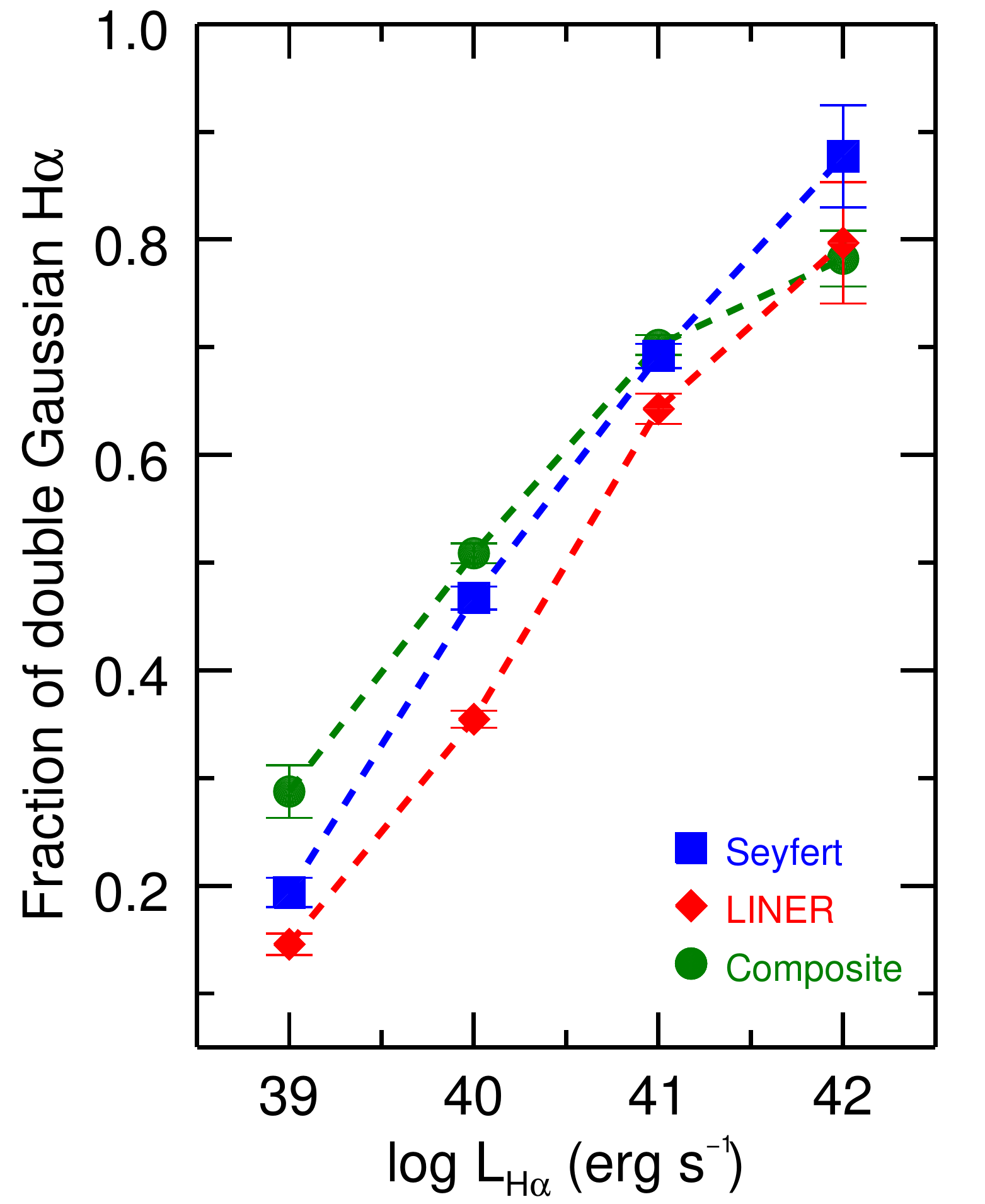}
	\caption{Fraction of galaxies with double Gaussian \ha\ profile as a function of \ha\ luminosity. We use poisson error as an error bar. Each color code represents groups : composite objects (green), LINERs (red) and Seyfert galaxies (blue).	\label{fig:fr1}}
\end{figure}

In this section, we investigate the outflow fraction based on the measured \ha\ kinematics. First, we examine the fraction of AGNs with double Gaussian \ha\ as a function of \ha\ luminosity. Since the double Gaussian profile (i.e., wing component) indicates the presence of outflow, we use this fraction as a proxy for the outflow fraction. Figure~\ref{fig:fr1} shows the increasing fraction with increasing \ha\ luminosity, from 20$\%$ to $\sim$90$\%$ over 4 orders of magnitude in \ha\ luminosity. All groups (Seyfert galaxies, composite objects, and LINERs) show a similar trend, as similarly found in the case of \oiii, while the overall fraction based on \ha\ is slightly lower than that of \oiii\ by 5-10\%. Note that the much lower outflow fraction at low luminosity should be taken as a lower limit since the detection of a wing component in \ha\ is presumably much more difficult for lower luminosity AGNs. 
		
\begin{figure}[t]
	\center
	\includegraphics[width=.49\textwidth, height=0.45\textwidth]{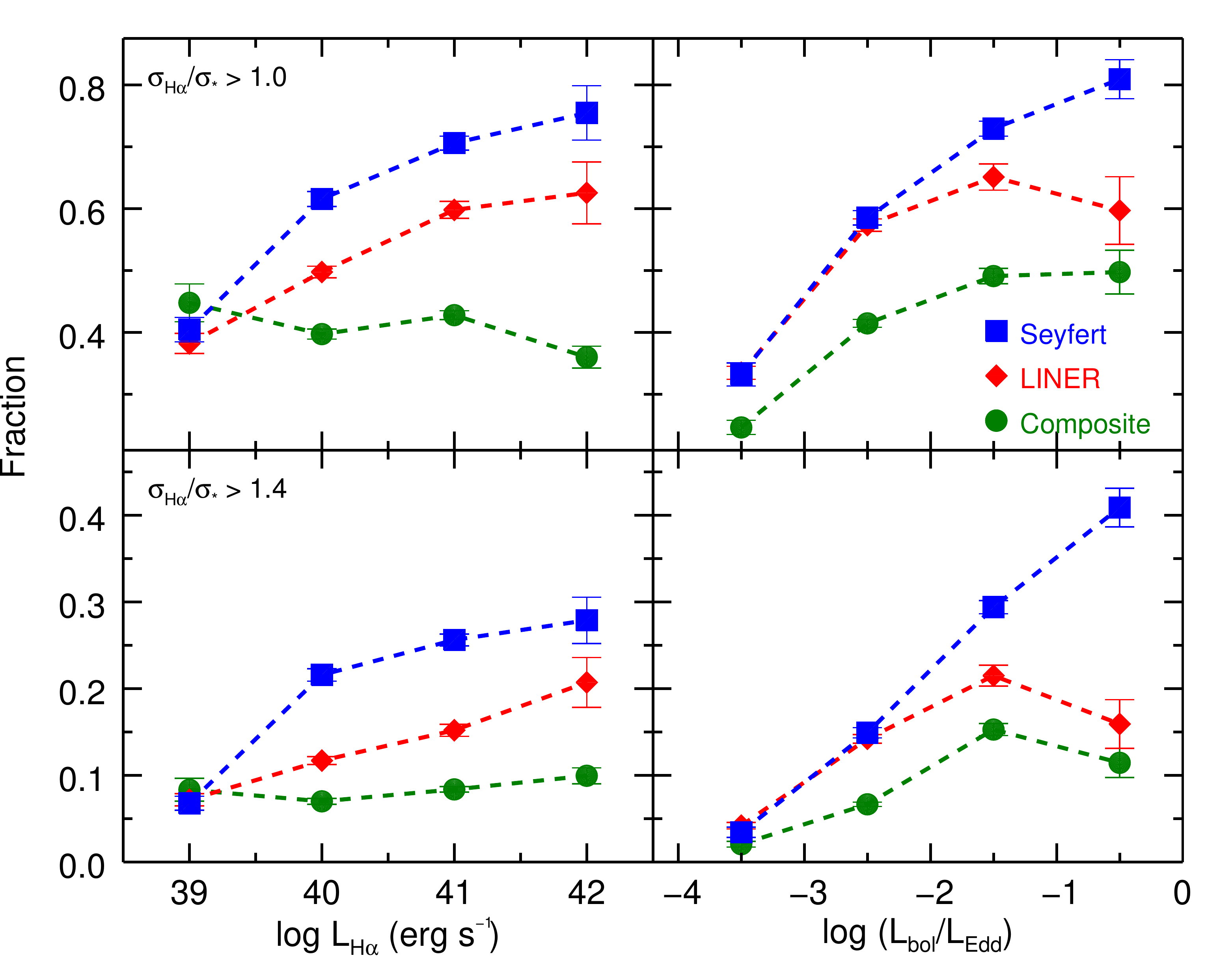}
	\caption{Fraction of galaxies with \ha\ to stellar velocity dispersion ratio above 1.0 (upper panel) and above 1.4 (lower panel) as a function of \ha\ luminosity (left) and Eddington ratio (right). The dispersion ratio value above 1.4 is where non-gravitational velocity dispersion is larger than gravitational velocity dispersion. Color-code is same as Figure~\ref{fig:fr1}. \label{fig:fr2}}
\end{figure}

Second, we use the measured \ha\ velocity dispersion to count AGNs with outflows. Since the observed total profile of \ha\ includes gravitational and non-gravitational (outflow) components (see Eq. 2), we assume that outflows are detected if  the measured velocity dispersion based on the total \ha\ profile is larger than stellar velocity dispersion (i.e., $\sigma_{H\alpha}$/$\sigma_{*}$ $>$= 1.0). For AGNs with strong outflows, we use a criterion that the outflow component in the \ha\ is comparable to or larger than the gravitational component, which is represented by stellar velocity dispersion (i.e., $\sigma_{H\alpha}$/$\sigma_{*}$ $>$= 1.4). Figure~\ref{fig:fr2} shows that the fraction of AGNs with detectable outflows increases from 40$\%$ to 80\% with increasing \ha\ luminosity in Seyfert galaxies. A similar trend is found in LINERs with overall smaller fractions while the outflow fraction in composite objects is $\sim$40\% without a significant change with \ha\ luminosity. When we count AGNs with strong outflows (i.e., $\sigma_{H\alpha}$/$\sigma_{*}$ > 1.4), we find a similar trend with \ha\ luminosity, but the fraction is much lower as expected. For Seyfert galaxies, the fraction is  25$\%$ at low luminosity and increases up to $\sim40\%$ at high luminosity, indicating that strong outflows are rare. LINERs shows a similar increasing trend with increasing \ha\ luminosity, but the fraction is lower than that of Seyfert galaxies at given luminosity. In the case of composite objects, we see that the outflow fraction is relatively flat over the luminosity range, implying that \ha\ line profile is significantly affected by the contribution from star-forming region.
		 		 
In summary we find that AGNs with a detectable kinematic signature in \ha\ are common among pure AGNs, particularly at high \ha\ luminosity and high Eddington ratio ranges, while the fraction of composite objects seems significantly different, presumably due to the contamination in the \ha\ line from star-forming region. 	 

\section{Discussion}
\label{sec:discussion}

\subsection{AGN outflow effect on \ha\ emission line}

	By comparing \ha\ velocity dispersion with stellar velocity dispersion, we find that hydrogen gas is not only tracing the gravitational potential of the host galaxy, but also influenced by the additional non-virial component (i.e., gas outflow). Outflow kinematics are often manifested by the wing component in \Ha\ (and also \oiii), which are common among AGNs with high \ha\ luminosity (80$\sim$90$\%$). Also, detectable outflows are more prevalent among Seyfert galaxies than among LINERs or composite objects. These trends indicates that the outflow component of \ha\ is related to AGN activity, implying that outflows are AGN-driven. 

	The amount of non gravitational effect exerted on \ha\ can be represented by the normalized \ha\ velocity dispersion by stellar velocity dispersion ($\sigma_{H\alpha}$/$\sigma_{*}$). By comparing this ratio with Eddington ratio, we find that outflow kinematics  increase with Eddington ratio albeit with large scatter, indicating the connection between outflow kinematics and AGN energetics. The fraction of AGNs with large dispersion ratios also demonstrates that strong outflow on \ha\ is common for \ha\ luminous AGNs or AGNs with high Eddington ratio. 
		
	As a tracer of gas outflows, \Ha\ velocity shift shows a correlation with AGN luminosity. However, the majority of AGNs has relatively small velocity shift less than 20~\kms, which is comparable to the measurement uncertainty. This is due to the intrinsic nature of Type 2 AGNs since the outflow direction is close to the plane of the sky, and the projected velocity measured from the line-of-sight is relatively small. 
	However, by counting AGNs with reliable velocity shift measurements, we find that the number ratio between blue shifted \Ha\ and redshifted \Ha\ increases with \ha\ luminosity, which supports the biconical outflow geometry combined with a dusty galaxy plane as in detail investigated by 3-D outflow models \cite{Bae2016}. The extinction due to the large scale dusty galaxy plane has to play a significant role in order to preferentially hide a part of the bicone so that the flux-weighted \Ha\ line is either blueshifted or redshifted
in the observed spectra (see the discussion in \cite{Bae2016}.
	
	The lower outflow fraction in AGNs with low luminosity or low Eddington ratio can be interpreted as the combination of two effects.
First, as we see the broad correlation between outflow kinematics and AGN luminosity, the kinematic signature of outflows, i.e., the wing component in \Ha\ or the velocity dispersion of \Ha, is much weaker in low luminosity AGNs. Consequently, it is more difficult to detect, 
leading to apparent lower outflow fraction. Second, for a given galaxy gravitational potential, a weak outflow signature (i.e., a wing) can be easily diluted by the virial motion, since the gravitational component will be dominating in shaping the \Ha\ line profile. This effect is more outstanding for composite objects since SF region can also substantially contribute to the observed \Ha. 

\subsection{Comparison of \ha\ kinematics with \oiii}

	In our previous study, \cite{Bae2014} compared velocity shift of \oiii\ and \ha\ using a subsample at z < 0.1, and demonstrated that \ha\ has smaller velocity shift compared to \oiii\ due to the contamination of star-forming region.  Although they used  single Gaussian models to fit \ha\ line and 
used the peak of the line to calculate velocity shift with respect to the systemic velocity, the reported results clearly indicated the difference of kinematics 
between \OIII\ and \ha. By consistently adopting double Gaussian models to \ha\ and \oiii, and using the 1st moment of the line profile in calculating velocity shift, our results presented in this paper supersedes that of \cite{Bae2014}, showing more consistent analysis compared with \OIII. 

	By comparing the kinematic measurements of \ha\ in this paper with that of \oiii\ presented in Paper~\Rnum{1}, we find that the outflow kinematics traced by \Ha\ is qualitatively similar to that of \oiii. However, the overall strength of outflows, i.e., velocity shift and velocity dispersion, and the fraction of outflows is relatively lower if we use the \ha\ line instead of \OIII. 
The difference is noticeable in the case of composite objects as expected since the contribution from non-AGN, i.e., SF region, is most significant compared to Seyfert galaxies and LINERs.
	
	While the limitation of this work is the lack of spatial resolution, \cite{Karou2016, Bae2017} used a subsample of AGNs with strong outflow signatures in \OIII\ to obtain integral field spectroscopy data. By applying the same kinematic analysis with double Gaussian models adopted in this paper to each pixel in the outflow region, they found that both narrow and broad components of \oiii\ are presenting non-viral outflow kinematics. 
	In the case of \ha, however, the broad component mainly reflects AGN outflows while the narrow component follows the stellar rotation due to the gravitation potential of the host galaxy. Also, they reported that the broad component of \ha\ at the center of the host galaxy is influenced by AGN, while the broad component of \ha\ detected in outer pixels is mainly representing star-forming region. 
These spatially resolved results suggest that it is more reliable to use the broad component of \ha\ measured from the central part of host galaxies for investigating the AGN outflow kinematics. In the case of \oiii, both narrow and broad components can be used for measuring AGN outflows. Without spatial resolution to separate AGN and star-forming regions, a careful interpretation has to be applied for the outflow analysis. Although there may be significant uncertainties of kinematic measurements for certain individual objects, our results based on a large sample provide statistical constraints on the outflow kinematics and fraction from the flux-weighted \ha\ line.

\section{Summary and Conclusion}
\label{sec:summary and conclusion}
	We used the spatially integrated spectra of $\sim$37,000 Type 2 AGNs at z < 0.3 to investigate the effect of AGN outflows on the \Ha\ line.  We compared the measured kinematics of \ha\ with those of \oiii\ and stars. The main results are summarized in this section.

\begin{enumerate}
\setlength{\itemsep}{1pt}  
\setlength{\parskip}{4pt}
\item By comparing \oiii\ and \ha\ luminosities, we find that \ha\ luminosity is significantly influenced by the contribution from star-forming regions, suggesting 
that \ha\ luminosity is not a good surrogate of AGN luminosity. 

\item \ha\ velocity dispersion has a non-linear correlation with stellar velocity dispersion, indicating the presence of a non-gravitational component (i.e., AGN driven outflow) in the \Ha\ line profile. 

\item The velocity shift and velocity dispersion of \Ha\ increase with AGN luminosity and Eddington ratio, suggesting that more energetic AGNs show stronger outflows. Among luminous AGNs, \ha\ tends to be more blue-shifted than red-shifted, which can be understood as a characteristic feature of biconical outflows in Type 2 AGNs.  

\item The fraction of AGNs with a wing component increases with \ha\ luminosity.  The fraction of galaxies with considerable non-gravitational component increases with \ha\ luminosity for pure AGNs, while the outflow fraction of composite objects shows a flat trend. The fraction also increases as a function of Eddington ratio, but the overall outflow fraction is smaller than that measured from \oiii\ in Paper~\Rnum{1}, indicating the outflow signature is weaker in \Ha.

\end{enumerate}

	Based on these results, we conclude that the \ha\ emission line is also strongly influenced by AGN driven outflow, even though the amount of the detected kinematic effect is relatively smaller than that of \oiii. Thus, \Ha\ line is a very useful tracer of AGN outflows if other high-ionization lines are not available. At the same time, a careful analysis needs to be done with \Ha\ since the contribution from SF region can be significant, particularly for composite objects.  

\acknowledgments 
	This work was supported by the National Research Foundation of Korea (NRF) to the Center of Galaxy Evolution Research (No. 2016R1A2B3011457).

\bibliographystyle{apj}

\begin{thebibliography}{}
\expandafter\ifx\csname natexlab\endcsname\relax\def\natexlab#1{#1}\fi

\bibitem[{Abazajian {et~al.}(2009)Abazajian, Adelman-McCarthy, Ag{\"{u}}eros,
  Allam, Prieto, An, Anderson, Anderson, Annis, Bahcall, Bailer-Jones,
  Barentine, Bassett, Becker, Beers, Bell, Belokurov, Berlind, Berman,
  Bernardi, Bickerton, Bizyaev, Blakeslee, Blanton, Bochanski, Boroski,
  Brewington, Brinchmann, Brinkmann, Brunner, Budav{\'{a}}ri, Carey, Carliles,
  Carr, Castander, Cinabro, Connolly, Csabai, Cunha, Czarapata, Davenport,
  de~Haas, Dilday, Doi, Eisenstein, Evans, Evans, Fan, Friedman, Frieman,
  Fukugita, G{\"{a}}nsicke, Gates, Gillespie, Gilmore, Gonzalez, Gonzalez,
  Grebel, Gunn, Gy{\"{o}}ry, Hall, Harding, Harris, Harvanek, Hawley, Hayes,
  Heckman, Hendry, Hennessy, Hindsley, Hoblitt, Hogan, Hogg, Holtzman, Hyde,
  Ichikawa, Ichikawa, Im, Ivezi{\'{c}}, Jester, Jiang, Johnson, Jorgensen,
  Juri{\'{c}}, Kent, Kessler, Kleinman, Knapp, Konishi, Kron, Krzesinski,
  Kuropatkin, Lampeitl, Lebedeva, Lee, Lee, Leger, L{\'{e}}pine, Li, Lima, Lin,
  Long, Loomis, Loveday, Lupton, Magnier, Malanushenko, Malanushenko,
  Mandelbaum, Margon, Marriner, Mart{\'{i}}nez-Delgado, Matsubara, McGehee,
  McKay, Meiksin, Morrison, Mullally, Munn, Murphy, Nash, Nebot, Neilsen,
  Newberg, Newman, Nichol, Nicinski, Nieto-Santisteban, Nitta, Okamura,
  Oravetz, Ostriker, Owen, Padmanabhan, Pan, Park, Pauls, Peoples, Percival,
  Pier, Pope, Pourbaix, Price, Purger, Quinn, Raddick, Fiorentin, Richards,
  Richmond, Riess, Rix, Rockosi, Sako, Schlegel, Schneider, Scholz, Schreiber,
  Schwope, Seljak, Sesar, Sheldon, Shimasaku, Sibley, Simmons, Sivarani, Smith,
  Smith, Smol{\v{c}}i{\'{c}}, Snedden, Stebbins, Steinmetz, Stoughton, Strauss,
  SubbaRao, Suto, Szalay, Szapudi, Szkody, Tanaka, Tegmark, Teodoro, Thakar,
  Tremonti, Tucker, Uomoto, {Vanden Berk}, Vandenberg, Vidrih, Vogeley, Voges,
  Vogt, Wadadekar, Watters, Weinberg, West, White, Wilhite, Wonders, Yanny,
  Yocum, York, Zehavi, Zibetti, \& Zucker}]{Aba2009}
Abazajian, K.~N., Adelman-McCarthy, J.~K., Ag{\"{u}}eros, M.~A., {et~al.} 2009,
  ApJS. Ser., 182, 543

\bibitem[{Bae \& Woo(2014)}]{Bae2014}
Bae, H.-J., \& Woo, J.-H. 2014, ApJ, 795, 30

\bibitem[{Bae \& Woo(2016)}]{Bae2016}
---. 2016, ApJ, 828, 97

\bibitem[{Bae {et~al.}(2017)Bae, Woo, Karouzos, Gallo, Flohic, Shen, \&
  Yoon}]{Bae2017}
Bae, H.-J., Woo, J.-H., Karouzos, M., {et~al.} 2017, ApJ, 837, 91

\bibitem[{Boroson(2005)}]{Boroson2005}
Boroson, T. 2005, AJ, 130, 381

\bibitem[{Ciotti \& Ostriker(2007)}]{Ciotti2007}
Ciotti, L., \& Ostriker, J.~P. 2007, ApJ, 665, 1038

\bibitem[{Crenshaw {et~al.}(2010)Crenshaw, Schmitt, Kraemer, Mushotzky, \&
  Dunn}]{Cren2010}
Crenshaw, D.~M., Schmitt, H.~R., Kraemer, S.~B., Mushotzky, R.~F., \& Dunn,
  J.~P. 2010, ApJ, 708, 419

\bibitem[{DeGraf {et~al.}(2015)DeGraf, {Di Matteo}, Treu, Feng, Woo, \&
  Park}]{DeGraf2014}
DeGraf, C., {Di Matteo}, T., Treu, T., {et~al.} 2015, MNRAS, 454, 913

\bibitem[{Dubois {et~al.}(2013)Dubois, Gavazzi, Peirani, \& Silk}]{Dubois2013}
Dubois, Y., Gavazzi, R., Peirani, S., \& Silk, J. 2013, MNRAS, 433, 3297

\bibitem[{Eun {et~al.}(2017)Eun, Woo, \& Bae}]{Eun2017}
Eun, D.-I., Woo, J.-H., \& Bae, H.-J. 2017, 842, 5

\bibitem[{Fabian(2012)}]{Fabian2012}
Fabian, A. 2012, ARA{\&}A, 50, 455

\bibitem[{Fabian {et~al.}(2006)Fabian, Celotti, \& Erlund}]{Fabian2006}
Fabian, A.~C., Celotti, A., \& Erlund, M.~C. 2006, MNRAS, 373, 16

\bibitem[{Garcia-Barreto {et~al.}(1996)Garcia-Barreto, Franco, \&
  Carrillo}]{Gar1996}
Garcia-Barreto, J.~A., Franco, J., \& Carrillo, R. 1996, ApJ, 469, 138

\bibitem[{Greene \& Ho(2005)}]{Greene2005}
Greene, J.~E., \& Ho, L.~C. 2005, ApJ, 627, 721

\bibitem[{Harrison {et~al.}(2014)Harrison, Alexander, Mullaney, \&
  Swinbank}]{Harrison2014}
Harrison, C.~M., Alexander, D.~M., Mullaney, J.~R., \& Swinbank, A.~M. 2014,
  MNRAS, 441, 3306

\bibitem[{Heckman {et~al.}(2004)Heckman, Kauffmann, Brinchmann, Charlot,
  Tremonti, \& White}]{Heck2004}
Heckman, T.~M., Kauffmann, G., Brinchmann, J., {et~al.} 2004, ApJ, 613, 109

\bibitem[{Karouzos {et~al.}(2016)Karouzos, Woo, \& Bae}]{Karou2016}
Karouzos, M., Woo, J.-H., \& Bae, H.-J. 2016, ApJ, 819, 148

\bibitem[{Kauffmann {et~al.}(2003)Kauffmann, Heckman, Tremonti, Brinchmann,
  Charlot, White, Ridgway, Brinkmann, Fukugita, Hall, Ivezi{\'{c}}, Richards,
  \& Schneider}]{Kauff2003}
Kauffmann, G., Heckman, T.~M., Tremonti, C., {et~al.} 2003, MNRAS, 346, 1055

\bibitem[{Kewley {et~al.}(2006)Kewley, Groves, Kauffmann, \&
  Heckman}]{Kewley2006}
Kewley, L.~J., Groves, B., Kauffmann, G., \& Heckman, T. 2006, MNRAS, 372, 961

\bibitem[{King \& Pounds(2015)}]{King2015}
King, A., \& Pounds, K. 2015, ARA{\&}A, 53, 115

\bibitem[{Kormendy \& Ho(2013)}]{Kormendy2013}
Kormendy, J., \& Ho, L.~C. 2013, ARA{\&}A, 51, 511

\bibitem[{Liu {et~al.}(2014)Liu, Zakamska, \& Greene}]{Liu2014}
Liu, G., Zakamska, N.~L., \& Greene, J.~E. 2014, MNRAS, 442, 1303

\bibitem[{Marconi \& Hunt(2003)}]{Marconi2003}
Marconi, A., \& Hunt, L.~K. 2003, ApJ, 589, L21

\bibitem[{Markwardt(2009)}]{Mark2009}
Markwardt, C.~B. 2009, ASPC, 411, 251

\bibitem[{Somerville {et~al.}(2008)Somerville, Hopkins, Cox, Robertson, \&
  Hernquist}]{Somer2008}
Somerville, R.~S., Hopkins, P.~F., Cox, T.~J., Robertson, B.~E., \& Hernquist,
  L. 2008, MNRAS, 391, 481

\bibitem[{Woo {et~al.}(2016)Woo, Bae, Son, \& Karouzos}]{Woo2016}
Woo, J.-H., Bae, H.-J., Son, D., \& Karouzos, M. 2016, ApJ, 817, 108 (Paper I)

\bibitem[{Woo {et~al.}(2014)Woo, Kim, Park, Bae, Kim, Lee, Kim, \&
  Kwon}]{Woo2014}
Woo, J.~H., Kim, J.~G., Park, D., {et~al.} 2014, JKAS, 47, 167

\bibitem[{Woo {et~al.}(2017)Woo, Son, \& Bae}]{Woo2017}
Woo, J.-H., Son, D., \& Bae, H.-J. 2017, arXiv:1702.06681

\bibitem[{Woo {et~al.}(2015)Woo, Yoon, Park, Park, \& Kim}]{Woo2015}
Woo, J.-H., Yoon, Y., Park, S., Park, D., \& Kim, S.~C. 2015, ApJ, 801, 38

\end{thebibliography}

\end{document}